\documentclass[11pt,a4paper]{article}
\usepackage[latin1]{inputenc}
\usepackage{amsfonts,amsmath,amstext,amssymb}
\usepackage[dvips]{graphicx}
\usepackage[english]{babel}

\newcommand{\llaves}[1]{\left\{#1\right\}}
\newcommand{\mult}[1]{\left\langle #1 \right\rangle}
\newcommand{\paren}[1]{\left(#1\right)}
\newcommand{\csplits}[1]{%
 \begin{equation*}\begin{split} #1
 \end{split}\end{equation*}}
\newcommand{\csplit}[1]{%
 \begin{equation}\begin{split} #1
 \end{split}\end{equation}}
\let\ds\displaystyle

\def\sn{\mathop{\rm sn }\nolimits}

\def\tn{\mathop{\rm tn }\nolimits}

\def\tanh{\mathop{\rm tanh}\nolimits}
\def\coth{\mathop{\rm coth}\nolimits}

\def\R{\mathbb{R}}

\def\a{\alpha}
\def\n{\nabla}
\def\g{\gamma}

\let\epsilon=\varepsilon

\def\g{\gamma}
\def\b{\beta}

\title{Geometrical particle models on $3D$ null curves%
\footnotetext{\hspace*{-17pt}${}^1$ aferr@um.es\\
$^2$ agpastor@um.es\\ $^3$ plucas@um.es, corresponding author. FAX number: +34-68-364182}}
\author{Angel Ferr\'andez$^1$, Angel Gim\'enez$^2$ and Pascual Lucas$^3$}
\date{\small Departamento de Matem\'aticas, Universidad de Murcia\\
 30100 Espinardo, Murcia, Spain}

\begin{document}
\maketitle

\begin{quote}
\footnotesize
\begin{center}
{\bf Abstract}
\end{center}
The simplest (2+1)-dimensional mechanical systems associated with light-like
curves, already studied by Nersessian and Ramos, are reconsidered. The action
is linear in the curvature of the particle path and the moduli spaces of
solutions are completely exhibited in 3-dimensional Minkowski background, even
when the action is not proportional to the pseudo-arc length of the trajectory.\\[2mm]
\emph{PACS numbers(s):} 04.20.-q, 02.40.-k\\[2mm]
\end{quote}

\section{Introduction}

There exists a very nice literature concerning Lagrangians describing spinning
particles, both massive and massless. It is well known that, in the general
situation, one has to provide the classical model with the extra bosonic
variables. An interesting hypothesis deals with Lagrangians on higher
geometrical invariants to supply those extra degrees of freedom.

The attractive point of view of this approach if that the spinning degrees of freedom
are encoded in the geometry of its world trajectories. The Poincaré and
invariance requirements imply that an admissible Lagrangian density $F$ must
depend on the extrinsic curvatures of curves in the background gravitational
field.

The mechanical systems depending on the first and second curvatures became
intensively studied in the late eighties as toy models of rigid strings
and (2+1)-dimensional field theories with the Chern-Simons term (see the work
by Polyakov). Before long, it became clear, mainly due to the studies of
Plyushchay, that those systems were of independent interest.

For instance, for $D=(2+1), F=a+bk_1+ck_2, a\neq 0$, they describe a massive
relativistic anyon, \cite{MR92e:81126,MR94k:81102}; for $D=(3+1), F=a+bk_1,
a\neq 0$, a massive relativistic boson, \cite{MR92a:81098}; for $D=(3+1),
F=ck_1$, a massless particle with an arbitrary (both integer and half-integer)
helicity, \cite{MR92c:81080}. In \cite{Barros:GRG2002} the author reconsiders
the simplest models describing spinning particles with rigidity, both massive
and massless, and describes the moduli spaces of solutions in (2+1)-backgrounds
with constant curvature. For $D=(1+1)$, the system with $F=a+bk_1^2$ corresponds
to the effective action of relativistic kink in the field of soliton,
\cite{kink-soliton}.

All actions considered before are defined on non-isotropic curves (spacelike or
timelike). But in a $(d+1)$-dimensional space one can study actions defined on
null (lightlike) curves.

In \cite{MR2000c:81113} the simplest geometrical particle model, associated
with null paths in four-dimensional Minkowski spacetime, is studied. The action
is proportional to the pseudo-arc of the particle. This geometrical particle
model provides us with an unified description of Dirac fermions ($s=1/2$) and
massive higher spin fields. In particular, the authors obtain the equations of
motion and show that they are particular examples of null helices. In
\cite{MR2000h:81357} the authors consider the same geometrical particle model
associated with null curves in (2+1)-dimensions. They show that under
quantization it yields the (2+1)-dimensional anyonic field equation
supplemented with a Majorana-like relation on mass and spin, i.e., mass
$\times$ spin = $c^2$, with $c$ the coupling constant in front of the action.

Unlike the known geometrical models of $4D$ spinning particles and $3D$ anyons,
the model \cite{MR2000c:81113} is formulated on light-like curves; but like the
known models with higher derivatives, the model has a spectrum similar to the spectrum
of the Majorana equation, which possesses the massive, massless and tachyonic
solutions. In \cite{PLB459:Plyushchay} it is shown that the model for massive
$4D$ spinning particles and $3D$ anyons with light-like worldlines may be also
constructed by reducing the model of spinning particles of a fixed mass to the
light-like curves.

In \cite{NPPS88:Nerssesian}, the authors consider a more complicated
three-dimensional system associated with null curves. The action is a linear
function in the curvature (sometimes called torsion) of the curve. The authors
show that its mass and spin spectra are defined by one-dimensional
nonrelativistic mechanics with a cubic potential. Consequently this system
possesses the typical properties of resonance-like particles.

In this paper we reconsider the above mechanical system and the motion
equations for these Lagrangians are rigorously obtained in (2+1)-background
gravitational fields.

The paper is organized as follows. In Section 2 we present the model, whose
action is given by
\[
S=2c\int(\lambda+\mu k)ds,
\]
where $\lambda$ and $\mu$ are constant, $s$ denotes the pseudo-arc length
parameter on the null curve and $k$ stands for its curvature. The motion
equations for these Lagrangians are completely given in (2+1)-background
gravitational fields. In Section 3 we solve the motion equations and obtain the
null worldlines of the relativistic particles. In Section 4 we sketch some
worldlines of the relativistic particles obtained in the preceding section, in
all studied cases. Section 5 is devoted to discussion and concluding remarks.

\section{The model and the equations of motion}

Let $M^3_1$ denote a 3-dimensional space-time with background gravitational
field $\mult{,}$, constant curvature $G$ and Levi-Civita connection $\nabla$.

We consider mechanical systems with Lagrangians which linearly depend on the
curvature of a light-like curve. This curvature function is sometimes called
torsion since it is obtained from the third derivative of the relativistic null
path. The space of elementary fields in this theory is the set $\Lambda$ of
null Cartan curves, \cite{FGL:NullHelices}, satisfying given first order
boundary data to drop out the boundary terms which appear when computing the
first order variation of the action.

Let $\gamma:I=[a,b]\to M^3_1$ be a null Cartan curve such that
$\llaves{\g'(s),\g''(s),\g'''(s)}$ is positively oriented for all $s\in I$ with
Cartan frame $\llaves{L,W,N}$, where $\mult{L,L}=\mult{N,N}=0$ and
$\mult{L,N}=-1$. The Cartan equations are given by (see \cite{FGL:NullHelices}
for details):
\begin{equation}
\label{EcuacionesCartan31}
\begin{split}
L'&=W, \\ W'&=-k L+N, \\ N'&=-kW,
\end{split}
\end{equation}
where the prime $()'$ denotes covariant derivative.

We consider the action $S:\Lambda\to\R$ given by
\[
S(\gamma)=2c\int_\gamma(\lambda+\mu k(s))ds.
\]
When $\lambda=1$ and $\mu=0$ it leads to the action studied by Nersessian and
Ramos in \cite{MR2000c:81113,MR2000h:81357}. The case $\mu=1$ has been
considered by Nersessian in \cite{NPPS88:Nerssesian}.

To compute the first-order variation of this action, along the elementary
fields space $\Lambda$, and so the field equations describing the dynamics of
this particle, we use a standard argument involving some integrations by parts.
Then by using the Cartan equations we have
\begin{equation}
\label{DerivaFuncional4} S'(0)=\left[\Omega\right]_a^b-c\int_a^b\mult{V,(\mu
k'''+3\mu kk'-\lambda k')L}ds
\end{equation}
where
 \csplit{ \label{Frontera} \Omega&=-c\mu L(h)+2c\mu
 \mult{\n_L^2V,N}+c (\mu k+\lambda)\mult{\n_LV,W} \\
 &\quad-c\mult{V,\n_L((\mu k+\lambda)W)} +2c(\frac{1}{2}\mu
 k''-\lambda k+2\mu G)\mult{V,L}+2c\mu k\mult{V,N}, }
$V$ standing for a generic variational vector field along $\gamma$ and
$h=-\mult{\n_L^2V,W}$.

We take curves with the same endpoints and having the same Cartan frame in
them, so that $\left[\Omega\right]_a^b$ vanishes. Under these conditions, the
first-order variation is
\[
S'(0)=-c\int_a^b\mult{V,(\mu k'''+3\mu kk'-\lambda k')L}ds,
\]
from which we obtain the following statement.\bigskip \clearpage

\begin{bf}
\noindent The trajectory $\gamma\in\Lambda$ is the null worldline of a
relativistic particle in the (2+1)-dimensional spacetime if and only if:
\vspace*{-\topskip}
\begin{enumerate}\itemsep0pt
\item[(i)] $W$, $N$ and $k$ are well defined on the whole world trajectory.
\item[(ii)] The following differential equation is satisfied
\begin{equation}
\label{EcuacionDiferencial} \mu k'''+3\mu k k'-\lambda k'=0.
\end{equation}
\end{enumerate}
\end{bf}

\section{The solutions of the equations of motion}

In this section we give a complete and explicit integration of the motion
equations of Lagrangians giving models for relativistic particles that linearly
involve the curvature of the null path.

We first observe that curves of constant curvature (i.e. null helices,
\cite{FGL:NullHelices,BonnorCurvas}) are always possible trajectories of the
particles, for any choice of $\lambda$ and $\mu$.

A second observation is that if the action is proportional to the pseudo-arc
length of the particle path (i.e. $\mu=0$ and $\lambda\neq 0$) then we have
that its solutions are also null helices. This was obtained in
\cite{MR2000h:81357}, where the authors show that the classical phase space of
this system agrees with that of a massive spinning particle of spin $s=c^2/m$,
where $m$ is the particle mass and $c$ is the coupling constant in front of the
action.

In what follows we analyze the case $\mu\neq 0$. Without loss of generality we
normalize the constant $\mu$ to be one.

A first integration of the equation gives us
\[
k''+\frac{3}{2}k^2-\lambda k+C=0,
\]
where $C$ is a constant. By standard techniques of integration, this equation
leads to
\begin{equation}\label{eq.dif}
(k')^2+k^3-\lambda k^2+2Ck+D=0,
\end{equation}
where $D$ is another constant. Note that constants $C$ and $D$ are not
arbitrary, since they are related with the mass $m$ and the spin $s$ of the
particle. In fact, when $G=0$ one can see that
\begin{eqnarray*}
2C &=& \ds\frac{ms}{c^2}-\lambda^2\\
D &=& \ds -\frac{m^2}{c^2}-\frac{ms}{c^2}\lambda+\lambda^3,
\end{eqnarray*}
so that
\[
D+2C\lambda=-\frac{m^2}{c^2}.
\]
If we assume that $k$ is constant, from (\ref{eq.dif}) we obtain that the
system has a local minimum, where the so-called ``semidiscrete''
(``semistationary'') or resonance-like levels can exist. The local minimun
(``ground state'') corresponds to the point $k=k_0$ defined by the equation
\[
3k_0^2-2\lambda k_0+\frac{ms}{c^2}-\lambda^2=0.
\]

Equation (\ref{eq.dif}) completely determines the geometry of the worldline, up
to congruences in the background gravitational field $(M^3_1,\mult{,})$. In
order to get the explicit solution of the motion equation, put $(k')^2=P(k)$,
where $P$ is a polynomial of degree 3. By using standard techniques involving
the elliptic Jacobi functions, the solution can be found according to the roots
$\alpha_1$, $\alpha_2$ and $\alpha_3$ ($\alpha_1\leq\alpha_2\leq\alpha_3$) of
the equation $P(t)=0$.

Before obtaining all the solutions, note that since $P(k)=(k')^2$ then $k$
takes values only where $P$ is non negative. Trivial solutions are
$k(s)=\alpha_i$, where $\alpha_i$ is a real root of $P$. Now we are going to
analyze all possible cases and present pictures of the corresponding curvature
functions.

\subsubsection*{\mathversion{bold}I. $P$ has a real root of multiplicity 3: $\a=\a_1=\a_2=\a_3$}

\begin{figure}[h]\centering
\includegraphics[width=5.89in,clip]{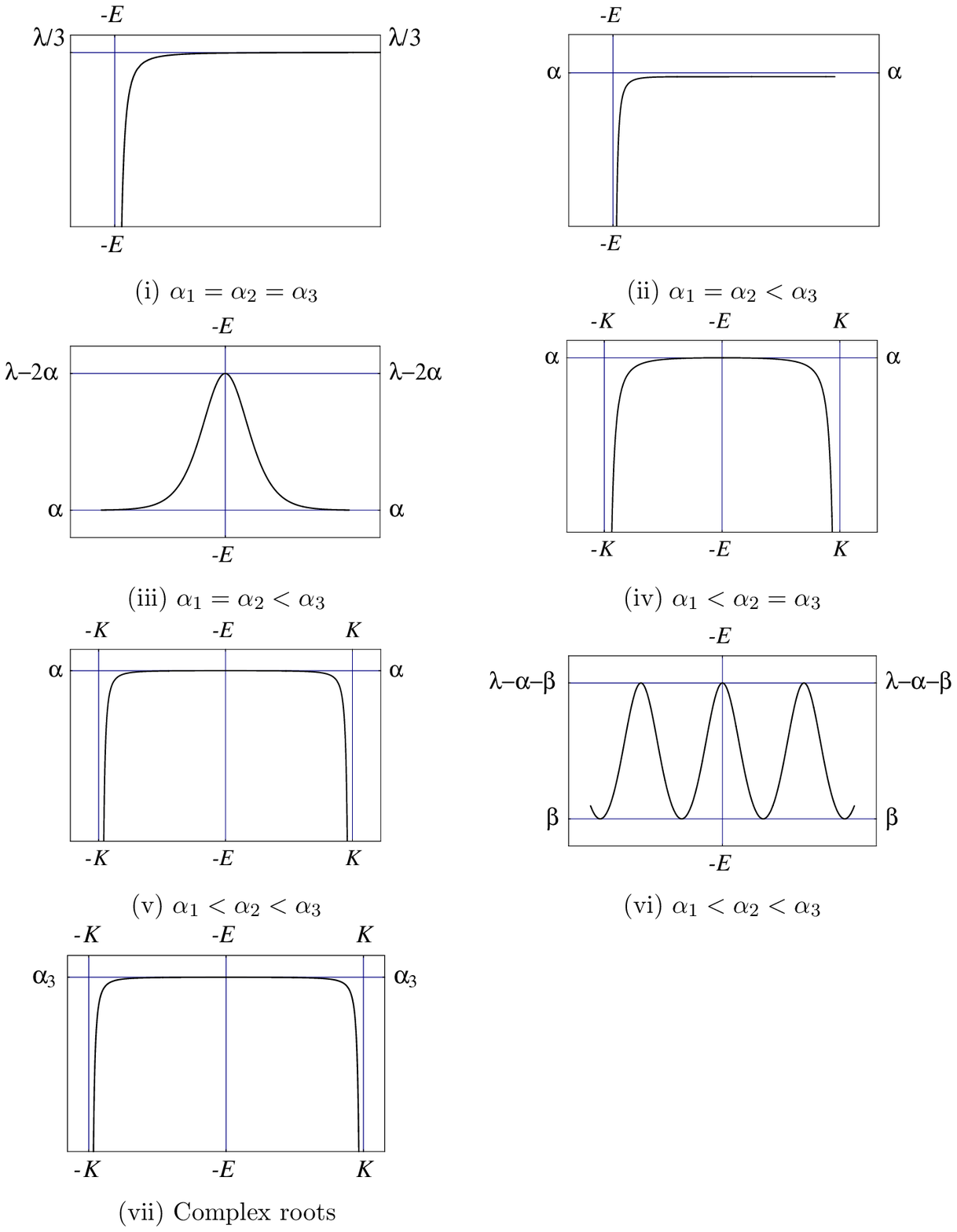}
\caption{\label{fig.curvatures}Curvature function for the different
possibilities of the roots of the polynomial $P$. It is quite interesting to
remark that in cases (i), (ii) and (iii), as $s$ increases, $k(s)$ approaches
to a constant, said otherwise, the trajectory looks like a helix.}
\end{figure}

We have that $\alpha=\lambda/3$ and the curvature function is given by
\[
k(s)=\frac{\lambda}3-\frac{4}{(s+E)^2},\qquad s\in (-\infty,\lambda/3)
\]
where $E$ is a constant of integration depending on the initial condition
satisfying that $s+E$ is always different from zero (see
Figure~\ref{fig.curvatures} (i)).

\subsubsection*{\mathversion{bold}II. $P$ has two real roots, the lowest with multiplicity 2:
$\a=\a_1=\a_2<\a_3$}

The root $\alpha_3$ is given by $\lambda-2\alpha$. There are two possibilities:
 \csplits{
 k(s)&=\lambda-2\a+(3\a-\lambda)\coth^2\paren{\frac{1}{2}\sqrt{\lambda-3\a}(s+E)},\quad s\in (-\infty,\a) \\
 k(s)&=\lambda-2\a+(3\a-\lambda)\tanh^2\paren{\frac{1}{2}\sqrt{\lambda-3\a}(s+E)},\quad
 s\in (\a,\lambda-2\alpha]
 }
where $E$ is a constant (see Figure~\ref{fig.curvatures} (ii)-(iii)).

\subsubsection*{\mathversion{bold}III. $P$ has two real roots, the greatest with multiplicity 2:
$\a=\a_1<\a_2=\a_3$}

We obtain that $\a_2=\a_3=(\lambda-\a)/2$, and the solution is given by
\[
k(s)=\a+\frac{3\a-\lambda}{2}\tan^2\paren{\frac{1}{2}
\sqrt{\frac{\lambda-3\a}{2}}(s+E)},\qquad s\in (-\infty,\a]
\]
where $E$ is a constant (see Figure~\ref{fig.curvatures} (iv)).

\subsubsection*{\mathversion{bold}IV. $P$ has three distinct real roots: $\a_1<\a_2<\a_3$}

Let us denote $\a=\a_1$ and $\b=\a_2$, then $\a_3=\lambda-\a-\b$. There are two
possibilities for the curvature:
 \csplits{
 k(s)&=\a-(\b-\a)\tn^2\paren{\frac{1}{2}\sqrt{\lambda-2\a-\b}(s+E),
 \sqrt{\frac{\lambda-\a-2\b}{\lambda-2\a-\b}}}, \\
 k(s)&=\lambda-\a-\b+(\a+2\b-\lambda)\sn^2\paren{\frac{1}{2}
 \sqrt{\lambda-2\a-\b}(s+E),\sqrt{\frac{\lambda-\a-2\b}{\lambda-2\a-\b}}},}
defined on the intervals $(-\infty,\a]$ or $[\b,\lambda-\a-\b]$, respectively
(see Figure~\ref{fig.curvatures} (v)-(vi)).

\subsubsection*{\mathversion{bold}V. $P$ has complex roots}

Let us suppose that $\a_1$ and $\a_2$ are complex (so $\a_3$ is real). Then the
curvature is given by
\[
k(s)=\a_3-(\a_3-\a_2)\sn^2\paren{\frac{1}{2}\sqrt{\a_3-\a_1}(s+E),
\sqrt{\frac{\a_2-\a_3}{\a_1-\a_3}}},\qquad s\in (-\infty,\a_3].
\]
(See Figure~\ref{fig.curvatures} (vii)).

\section{\mathversion{bold}The worldlines of the relativistic particles in $3D$ Minkowski spacetime}

\begin{figure}[h]\centering
\includegraphics[width=4.77in,clip]{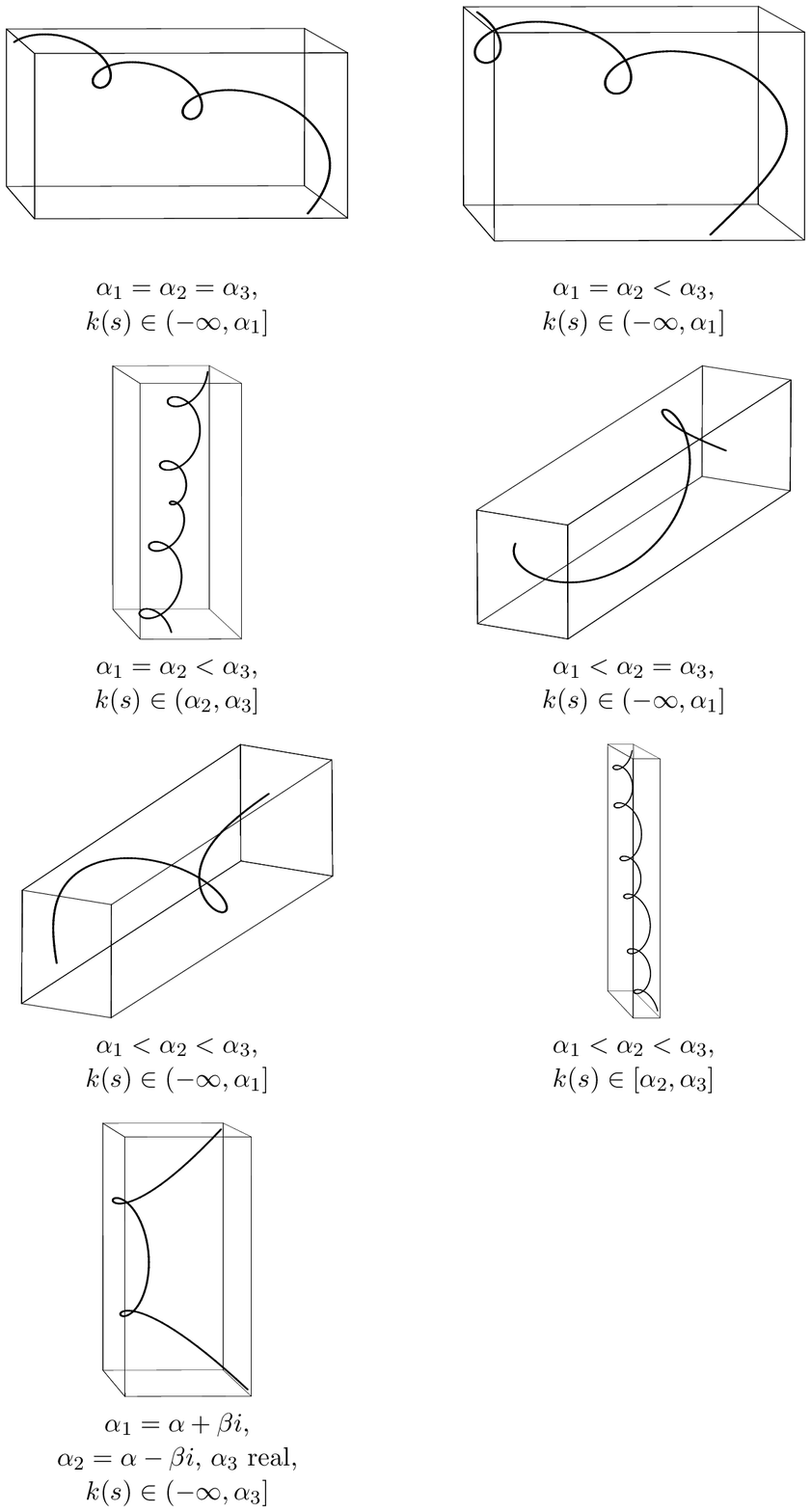}
\caption{\label{fig.worldlines}Worldlines for different curvature functions}
\end{figure}

Once we know the curvature functions, the worldlines of the relativistic
particles can be obtained by integrating the Cartan equations. The explicit
integration of these equations is a difficult task, sometimes impossible (even
when the curvature is a nice function). In our case, the goal of finding the
exact worldlines can be reached by numeric integration. In Figure
\ref{fig.worldlines} we sketch (with the help of
\textsc{Mathematica}) the particle worldlines in all discussed
cases in the preceding section.

\section{Discussion and outlook}

We have studied the action whose Lagrangian is linear in the curvature of the
particle path, completing previous work by Nersessian and Ramos. Our results,
following a totally different procedure, agree with those of the above
mentioned authors. For example, when the action is proportional to the
pseudo-arc length of the particle path, we obtain that the trajectories of the
relativistic particles of the model are null helices. In the general case, when
the curvature appears linearly in the action, the trajectories are
characterized by equation (\ref{eq.dif}). This equation is essentially
equivalent to equation (39) of \cite{NPPS88:Nerssesian}, where the authors were
not able to obtain the trajectories of the particles. Here, and using
geometrical methods, we can obtain a complete description of the relativistic
particle paths.

To conclude, let us indicate some problems that deserve further attention.

First, even though we have got an explicit description of the motion equation at
$D=(2+1)$ with constant curvature, we note that a priori there is no restriction to apply these ideas in
background gravitational fields with non-constant curvature. Then we can study what are the
trajectories of the relativistic particles in this situation.

Secondly, in the non-isotropic case the system with $F=a+bk_1^2$, where $k_1$
is the curvature of the worldline trajectory, corresponds to the effective
action of relativistic kink in the field of soliton, \cite{kink-soliton}. As a
generalization of our model we could consider an action whose Lagrangian $F$
depends quadratically on the curvature. Then, what are the equations of motion
of this action? Is it possible to integrate them? In this case, can we find all
the trajectories of the relativistic particles of the model?

\section*{Acknowledgements}

This research has been partially supported by Direcci\'on General de Investigaci\'on (MCYT)
grant BFM2001-2871 with FEDER funds. The second author is supported by a FPPI Grant, Program PG,
Ministerio de Educaci\'on y Ciencia.

\end{document}